# A Pacemaker with P=2.48 hour Modulated the Generator of Flares in the X-ray Light Curve of Sgr A* in the year 2012

Elia Leibowitz

School of Physics & Astronomy and Wise Observatory
Sachler Faculty of Exact Sciences
Tel Aviv University

elia@wise.tau.ac.il

**Abstract**

In an intensive observational campaign in the 9 month duration of *Chandra* X-ray Visionary Project that was conducted in the year 2012, 39 large X-ray flares of Sgr A* were recorded. An analysis of the times of the observed flares reveals that the 39 flares are separated in time by intervals that are grouped around integer numbers times 0.10333 days. This time interval is thus the period of a uniform grid of equally spaced points on the time axis. The grouping of the flares around tic marks of this grid is derived from the data with at least a 3.2 $\sigma$ level of statistical significance. No signal of any period can be found among 22 flares recorded by *Chandra* in the years 2013-2014. If the 0.10333 d period is that of a nearly circular Keplerian orbit around the blackhole at the center of the Galaxy, its radius is at 7.6 Schwarzschild radii. Large flares were more likely to be triggered when the agent responsible for their outbursts was near the peri-center phase of its slightly eccentric orbit.

# 1. Introduction

A characteristic feature of the X-ray luminosity of Sgr A*, the supermassive blackhole (SMBH) at the center of the Galaxy, is the occurrence of flares at a rate that is roughly one per day. A list of references to early reports on flare detections, as well as of various interpretations that have been suggested for them, is provided by Li et al (2015). The most extensive observations of this phenomenon were conducted by the $Chandra\ X-ray\ observatory$ in the X-ray Visionary Project (XVP) performed between February and October 2012 (Neilsen et al, 2013 - henceforth N13). The object was observed 38 times along this period, with exposure times ranging from 14.53 ks to 189.25 ks.

N13 report on 39 flares that have been recorded during XVP, and Table 1 in their paper presents some of the basic parameters that characterize them, These include the beginning and the end times, the fluences and the peak count rates, as well as some other parameters of each flare. These authors analyzed the distributions of the fluences, durations, peak count rates and luminosities of these XVP detected flares. They further investigated the flux distribution of the X-ray light curves and suggested that flares may be the only source of X-ray emission from the inner accretion flow onto the BH at the center.

One of the characteristics of the SMBH Sgr A* that makes it particularly interesting is the ability to detect signals from an ever decreasing distances from the event horizon of this object. One can hardly overestimate the importance of recording signals from the close vicinity of black holes, of stellar masses and of SMBH, for their implications on the behavior of spacetime in regions of a very strong gravitational field, on the physics of blackholes in general and on the dynamics and physical properties of matter under extreme physical conditions.

Li et el (2015) performed a further detailed statistical analysis of the XVP data, which revealed some statistical properties and correlations among some of the measured flare parameters. Their analysis enabled these authors to create a model X-ray light curve of the source, with statistical characteristics similar to the observed ones. However, as stated by Li et al, the nature and origin of Sge A* flares are still under debate.

In this paper I want to draw attention to another statistical property of the 39 detected flares that may shed some more light on the origin and nature of this phenomenon. I show with a high level of statistical confidence that these flares occurred preferably at or close to such points on the time axis that the intervals between them are multiples of integers and one fixed time duration.

More recently a new list of 92 flares recorded between the years 2000 and 2014 was published by Ponti et al (2015 - henceforth P15). This list includes the 2012 observations, although with somewhat different partitions between the individual events. The same periodicity is found also among the 2012 events of this new list, however no period can be found in the 2013 and the 2014 data. The data of previous years are very thinly spread in time, some 24 flares in the period of 11 years from 2001 to 2011. This is hardly an appropriate data base for conducting a search for a periodicity of the order of a fraction of a day.

## 2. Analysis

In this section I consider the 39 flares as presented in N13. We are interested in the question whether or not there is some regularity in the distribution of the flare events along the time axis, beyond the mean frequency in their occurrences of about 1 or 1.1 flares per day, as is by now quite well established (Markoff (2010), Li et al (2015), N13).

In the published table of data, each flare comes with two specific times, the beginning and the end times. As seen in Table 1 of N13, flares vary considerably in their duration, between the extremes of 400 and 7800 seconds. In order to study the timing of the recorded flare events one has to assign to each flare a unique time value that will represent the occurrence time of that event. Naturally, this event time should be some well defined function of the beginning and of the end times of the flare, or of some other feature characteristic of this flare. The midpoint between the beginning and the end times naturally suggests itself as an appropriate representation of the time of the event. Figure 1 presents, in the y direction, the times t(i) of all i=1 to 39 events so defined. Times are MJD-55960 and are expressed in fractions of a day. I refer to this list of times as the m-series.

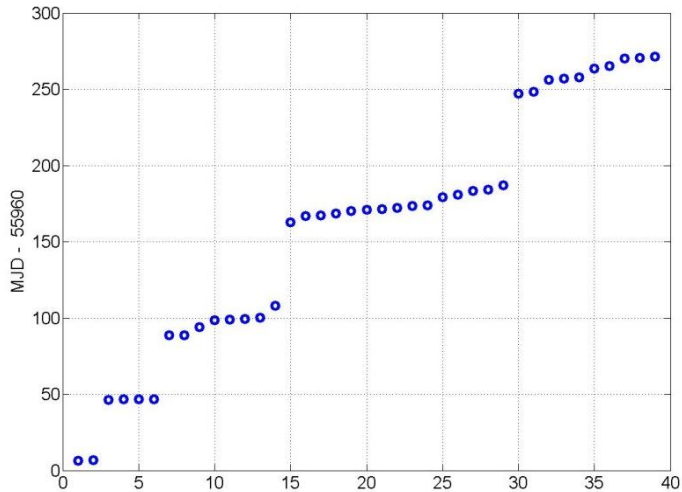

Figure 1: Times of the mid-points of 39 flares detected in the X-ray light curve of Sgr A* in 2012,

### 2.1 Power Spectrum

As a first step in a search for some regularity in the times of flare outbursts I constructed a time series [t,y] to represent the 2012 X-ray light curve (LC) of Sgr A* as observed by *Chandra*, as follows. The data presented in Table 1 of N13 were extracted from a LC binned into equal size bins of 300 sec width. The table presents the beginning time of each one of the 38 runs of the telescope, as well as the corresponding exposure times. Accordingly each time slot of the $Chandra$ observational window was covered by a grid of equi-distance points with 300 sec interval between them. Each of the 10797 points so established was assigned the value 0 as its y value. The times of the 39 mid-points of the observed flares were added to this "quiescence" LC as spikes with 1 as y value.

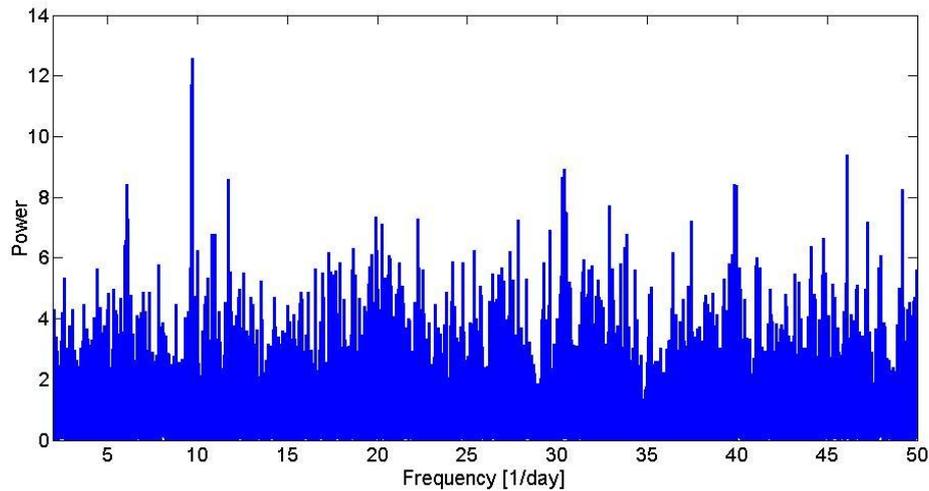

Figure 2: Power spectrum of an artificial LC of spikes, representing the location of the mid-points of the 39 flares in the 2012 observational window of the $Chandra$ X-ray telescope that has recorded them.

Figure 2 is the power spectrum of this time series in the frequency interval corresponding to the period interval [0.02-0.5] day. The highest peak in the eriodogram is centered at the frequency 9.6783 $d^{-1}$, corresponding to a period P=0.10332 days. Using expression (18) in Scargle (1982), the false alarm probability (FAP) of this peak may be estimated as 0.12.

The presentation of the above PS is not intended to be evidence of statistical significance for some regularity in the timing of the flares. It serves only as a qualitative hint to the possible existence of some regularity in the data. Indeed, the power spectrum technique is a poor method for looking for a periodic signal in a time series of the type presented here. The algorithm is trying to fit the harmonic waves even to all sections which have no flare in them. In fact, when I create a more realistic LC from the data presented in Table 1 of N13, taking into account also the different durations of the flares, as well as the large differences between their peak count rates, the PS of the resulting LC shows a general rise towards lower frequencies, with no notable single peak. A very weak signal as a low peak, hardly outstanding above the noise, can be identified at p=0.103d in the corresponding periodogram. I therefore turn to another search technique more suitable for our purpose.

## 2.2 Search of periodic grid

The non-existence of a significant signal in the PS of the m-series suggests that in this time series there is hardly a strictly periodic content in the [0.02-0.5] search interval. During the duration of the XVP project, there are a number of time intervals, longer than 0.5 d, in which no outburst of the object has been recorded. There is however the possibility that while the flare-generating mechanism is not necessarily periodic, it may still be governed by some process that makes the outbursts of the source to occur more likely around time points that form a periodic grid on the time axis. For a given set of events that occur at times t(i), measured starting from a common reference time t(0), that have this property, there is a period p, such that each event occurs at or close to a certain point k on an equally spaced grid on the time axis, whose distance from t(0) is $r_k p$, where $r_k$ is an integer.

In a search for such a significant period for our set of the N=39 events, if one exists, I consider 40000 p values, the frequencies of which are equally spaced in the period interval [0.02-0.5] d. For each of the N t(i) values, I calculate for each p(j) within the period search interval, its distance from the nearest point of the $r \times p(j)$ grid, as a fraction of the value of the period p(j). In other words, I find the distance of t(i) from the nearest grid point that is equal or smaller than p(j)/2. I therefore define

(1) $$f[i, q(j), t(0)] = dif\left[\frac{t(i)-t(0)}{p(j)}\right]$$

Here $dif$ is the decimal fraction $f$ of the distance of t(i), expressed in units of p(j), from its nearest integer number : $-\frac{1}{2} \leq f \leq \frac{1}{2}$.

For each value of p(j) I compute the variance of the ensemble of the $f$ values corresponding to the given N t(i) values:

(2) $$s^2[p(j), t(0)] = Var(f[i, p(j), t(0)])$$

For each of the p(j) values, the value of $s^2[p(j), t(0)]$ is computed 2n+1 times with n=5 different values of t(0). They are n equally spaced points on the time axis that cover half a cycle of the period p(j) on each side of some reference time value. The variance of the observations with respect to the tested period p(j), $S^2[p(j)]$, is the smallest value of $s^2[p(j), t0]$ among those obtained for all the t(0) parameter values considered. The value p(j) that minimizes $S^2[p(j)]$ is the best period within the search interval and the corresponding t(0) value is taken as an initial, reference time, associated with this period.

If a series of events harbors precisely a certain period p, i.e. if all events are separated from one another by integer numbers of this period, the standard deviation S(p) should be equal to 0.

If the events under consideration are preferably grouped around tic marks of a grid on the time axis of a periodicity p, the value of $S(p)$ will be smaller than $\sqrt{\frac{1}{12}}$, the StD of a rectangular distribution of numbers over the [-1/2,1/2] phase interval.

In general S(p) is a measure of the dispersion of the set, i.e. of how closely grouped are the observed set of events around the tic marks of the grid with the equal spacing of period p. The smaller the S value, the tighter is the grouping of the given series.

Figure 3a is a plot of the S(p) values for all the 40000 tested periods p, computed for the series of the 39 m-points, that have the smallest value among the $n$ =5 values of t(0) on each side of some reference time. There is a clearly distinguishable sharp minimum value of S at the frequency 9.6776 $d^{-1}$, corresponding to the period value P=0.10333 day. One reference time of the periodic grid on the time axis is t(0)=55966.0. The corresponding value of the S parameter is $S_m$ =0.16557. From the width of the spectral feature of the minimum corresponding to the period P we derive an uncertainty of 0.0001 $d \approx 10\ sec$ in the value of P.

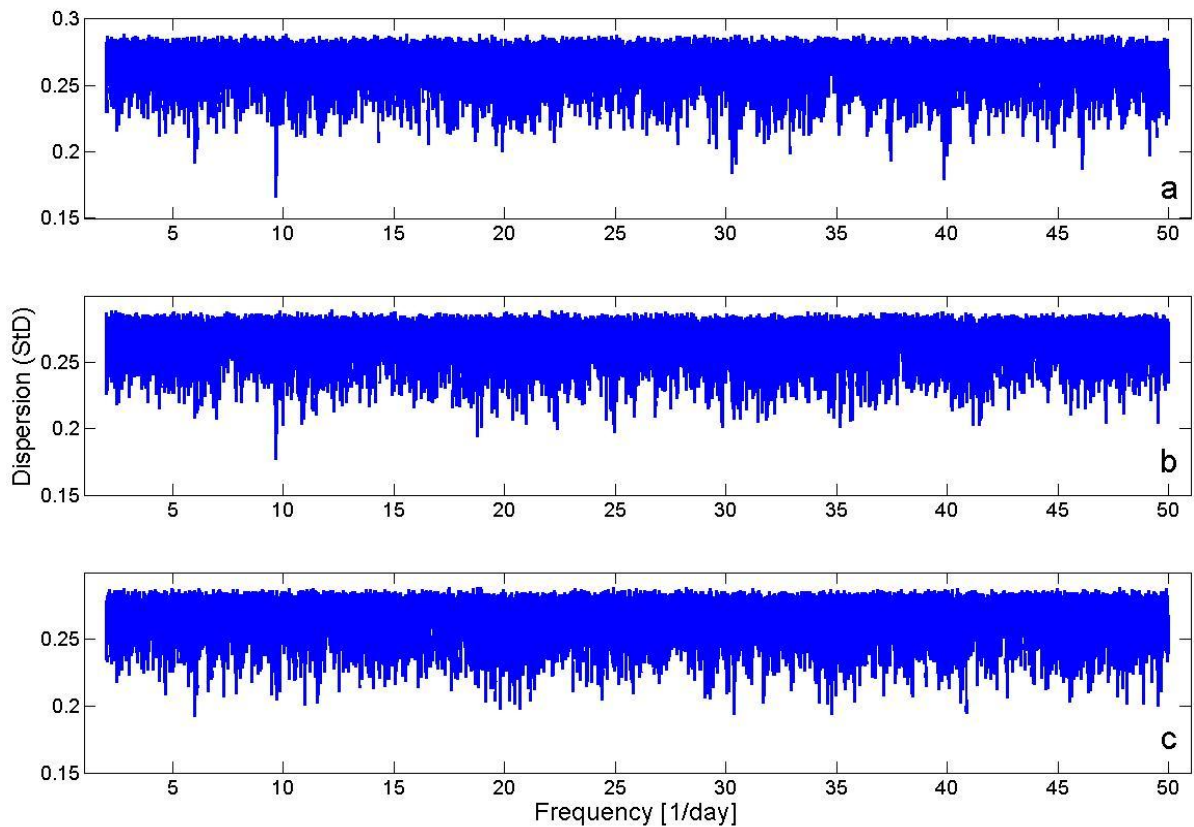

Figure 3: The StD parameter S, computed for 40000 period values in the range [0.02-0.5] d. (a) For the times of the mid-points of the 39 2012 flares of Sgr A*. (b) For the series of the beginning times of the flares. (c) For the end times of the flares.

Figure 3b is similar to frame a, where the computations were done with respect to the series of the beginning times of the flares, as given in Table 1 of N13. We call this set of events the b-series. Frame c is the same for the series of the end times of the flares, to which we refer as the e-series.

Comparing the periodograms of the 3 frames one notices that the minimum at the frequency 9.6776 $d^{-1}$, corresponding to P=0.10333 d in frame a, $S_m$=0.16557, is markedly lower than the one in frame b, $S_b$= 0.17666. In frame c the minimum feature at this frequency is being lost entirely in the noise.

The distribution of the time points of the m-series is of course dependent on those of the two other series. As already mentioned above, according to Table 1 of N13, the durations of the flares vary in an apparent non systematic manner between 400 and 7800 seconds, 0.005-0.09 d. This range is of the order of the period that we identify in the m-series, and to a somewhat lesser confidence level also in the b-series. The apparent lack of correlation between intervals of the order of P between the b-series and the e-series explains well the disappearance of the P minimum in frame c.

## 3. Significance

### 3.1 Test I

The minimum value of $S_m$=0.16557 is much smaller than the dispersion of a rectangular distribution of points over the [-1/2,1/2] phase interval. This inplies that the 39 midpoints of the 39 flares of N13 are indeed grouped rather tightly around tic marks of the p grid on the time axis. This result is obtained for a p value that was found by the search routine described in the previous section. In order to evaluate the significance of our finding for the measured m-series, we need to estimate the false alarm probability (FAP) of our finding. This is the probability that applying our search procedure on 39 time points that are chosen randomly, as described below, will yield a periodicity with a dispersion S of the 39 points that is equal or smaller than the $S_m$ value found for the real data.

For this purpose I performed a bootstrap type test (Efron and Tibshirani, 1993) by applying the analysis described in the previous section on a sample of 13878 pseudo-observed series of 39 numbers, simulating the observed m-series as follows. A pseudo-observed set is the set of the m-series, in which I replaced the decimal fraction of the day at which each of an m-series point occurred, with a random number from a rectangular probability distribution over the [0,1) interval. A pseudo observed set is thus preserving all the temporal content of the real data, except for the hour in the day at which each corresponding real event was recorded.

It should perhaps be noted that with pseudo-observed events consisting of series of 39 points selected randomly over the entire time interval of the XVP project, the resulting FAP is much smaller than the one obtained with our bootstrap sample, as indeed is expected from simple considerations. The search routine was applied on each set of the pseudo-observed events as defined above, with the same 40000 tested p values and $n = 5$ as was done for the real data, searching over the same search interval of [0.02-0.5] d for the period that is represented by the lowest minimum in the periodogram corresponding to Figure 3.

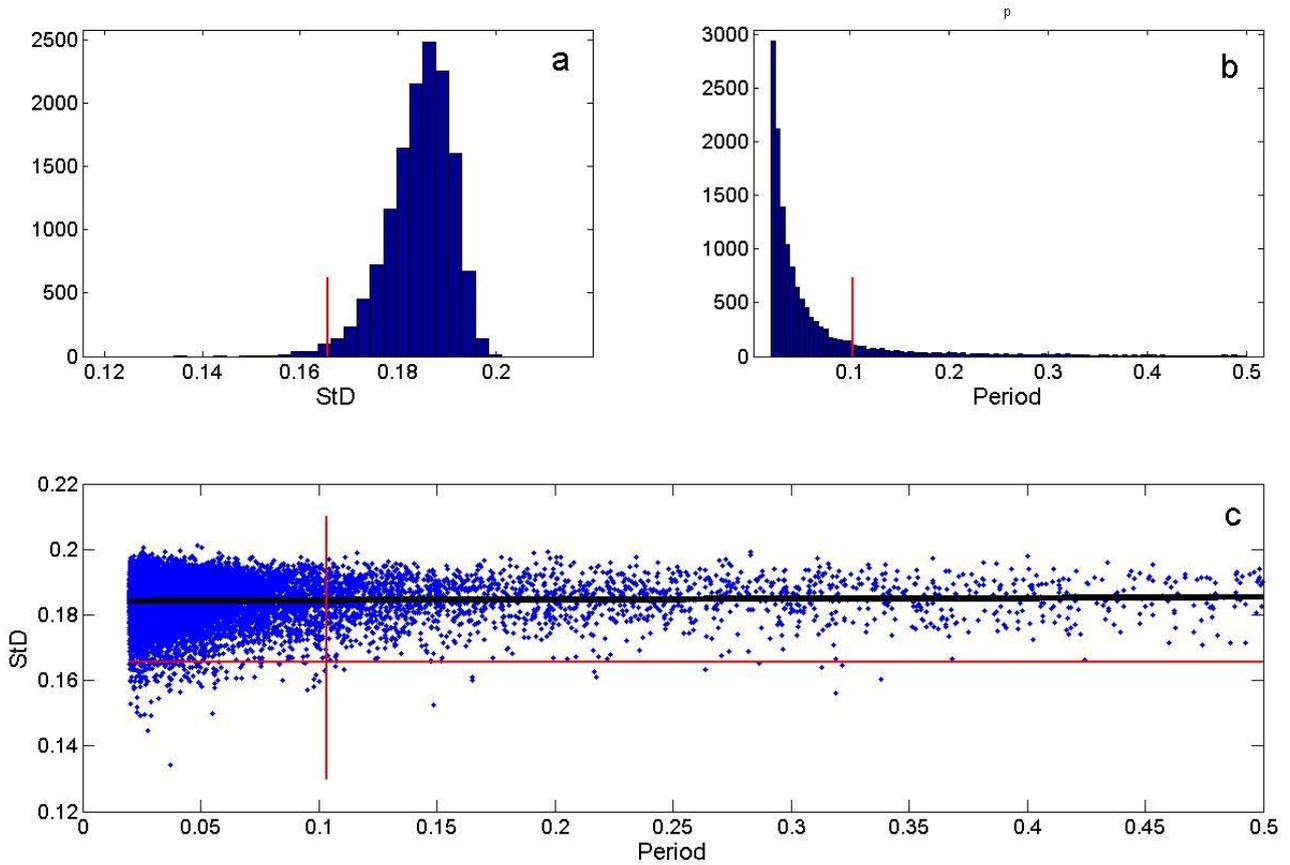

Figure 4: (a) Histogram of the S values found for each one of 13878 sets of 39 pseudo-observed flare events occurring at random hours in the days of the real observations. Vertical line marks the value of S found for the measured m-series. (b) Histogram of the corresponding period values. Vertical line marks the p value of the m-series. (c) The dispersion (StD) S value vs. the corresponding p values of the sample sets. The cross indicates the coordinates $(P, S)$ found in the real data. Horizontal line at the center of the distribution of the points is the linear regression line of S over p. See text for further explanation.

Figure 4a is the histogram of the S values found in a set of 13878 series of pseudo events so constructed. The vertical line marks the value $S_m$=0.16557 on the S axis. There are 166 points to the left of this line. Figure 4b is the histogram of the p values corresponding to the S values of frame a. Here the vertical line marks the value P=0.10333 on the p axis, with 1636 points on its right side. Figure 4c is a plot of the S values found in this sample of the 13878 series of 39 time points vs. the corresponding p values. The cross is positioned at the coordinates (P,$S_m$). The horizontal line at the center of the plot is the regression line of the S over the p values. The correlation coefficient between these two parameters in the sample is 0.024. It indicates that there is no correlation between the S and the p values of these series. Therefore the probability of an S value of a series of pseudo-observed points to be on the left side of the vertical line in frame a is independent of the probability of its p value to be on the right side of the vertical line in frame b.

The probability of a pseudo-observed series to harbor a period p≥P with a corresponding s≤ $S_m$ is therefore $FAP_1 = \frac{166*1636}{13878^2} = 0.00141$. In terms of StD units this $FAP_1$ value corresponds to a statistical significance within $3.19\sigma$. The independence of the S and p parameters implies also that the FAP can be obtained from the sample by the fraction of points of the sample that occupy the lower right side of the cross in Figure 4c, i.e. the fraction of the points for which $s \leq S_i$, and for which, at the same time, $p \geq P$. As can be seen in Figure 4c, this number is 13 and the FAP implied by this number is $FAP_2 = \frac{13}{13878} = 0.0009367$ corresponding to statistical significance within $3.31\sigma$. The difference between the 2 estimates is due to the finite size of the sample. We may conclude this section by asserting that the period P=0.10333 with the dispersion $S_m = 0.16557$ as found in the real data is statistically significant at a $3.2\sigma$ confidence level.

## 3.2 Further refinement

Figure 3 shows that for the m-series, the points of which are the middle of the b- and the e-series, we obtain a periodicity that fits this series better than for the b-series, and certainly better than for the e-series. This indicates that for each flare, somewhere in between the beginning and the end times, there is a time point such that the series consisting of all these points is grouped around points of the p grid on the time axis tighter than the beginning or the end times. The midpoints of the flares that were chosen as a natural guess are just one set of such points but not necessarily the best ones. In order to look for possible better definition of flare time points, that must of course be defined a-priori, before considering any period for the data, and must depend on the beginning and the end times of each individual flare, uniformly for all flares, I consider the following general formula for the time of a flare:

(3) $$t(a) = t_b + a \times (t_e - t_b) .$$

Here $t_b$ and $t_e$ are the beginning and the end times of the flare and $0 \leq a \leq 1$ is a free parameter. The value of $a$ determines the relative position of the point considered *The* time of the flare, along the duration of the flare. It takes of course the same value for all 39 observed flares. The points considered in the analysis so far are those with $a = 0.5$.

I now perform the search routine in a 3 dimensional space whose coordinates are the 40000 frequencies within the period search interval [0.02,0.5], the number of initial times $n = 5$, taken on each side of an arbitrary zero time, and with $n_a = 6$ different $a$ values, equally spaced between 0 and 1. I find that with $a$=0.4, the 39 times are grouped around the P grid points even tighter than the midpoints, with the fitting parameter value $S_I = 0.16222$. I call this set of 39 points the I-series.

## 3.3. Test II

It is worthwhile to underline here again that in the generation of the set of points to be analyzed, the existence of a periodicity in the data never enters the definition of the points, and in the search routine there is no a-priori assumption about an expected value of the period, other than in the definition of the search interval. In particular, all definitions of the time point representing a flare rest entirely on the beginning and the end times of that individual flare, with the same formula and the same $a$ value employed on all flares in a set, independently of the times of all other flares and with no reference to any possible period in the set.

Here again, I have estimated by numerical simulation the false alarm probability to find an equi-distance grid on the time axis of period P=0.10333 d, with the corresponding $S_I = 0.16222$ value. For this purpose I created sets of pseudo flares, rather than sets of pseudo-flare points as in the previous test. Each of the points generated by the previous simulation is now considered a beginning time of a pseudo-observed flare. I attach to each such beginning time a time segment, the length of which is selected randomly, with repetitions, from the ensemble of the 39 lengths of the observed flares. On each one of these 13878 sets of pseudo-observed flares I now apply the same 3 dimensional search routine as the one applied on the real data. In particular, for each set in this sample the search routine was applied 6 times for each value of the parameter $a$. The series of points to be compared with that of the I-series are those that belong in each set to the $a$ value that yields the smallest S value.

The histograms and the plot of S vs. p of the 13878 series so created look very much the same as in Figure 4. There is also no correlation between the S and the p parameters values in this set of series. As in the previous test we obtain here that $FAP_1 = 0.00109$, corresponding to $3.27\sigma$, and $FAP_2 == 0.00151$, corresponding to $3.17\sigma$.

There is also no appreciable dependence between the S and the p values in the set of all $6 \times 13878 = 83268$ series of 39 pseudo-events that we have considered. The formal correlation coefficient between these 2 parameters is 0.017042. In this ensemble of simulated sets we find 493 sets with $S \leq 0.16222$ and 11275 sets with $p \geq 0.10333\ d$. This gives $FAP1 = \frac{493 \times 11257}{83268^2} = 0.0008017$, corresponding to a significant level of $3.35\sigma$. In the same sample we find 68 sets with $S \leq 0.16222\ and\ p \geq 0.10333\ d$, yielding $FAP2 = \frac{68}{83268} = 0.0008166$, corresponding to the same confidence level of $3.35\sigma$.

The results of the above two tests allow us also to reject the hypothesis that the period P found in the data is in fact a result of some unknown effect of the time intervals of the order of a day between flares, on the exact timing of the flares within the day of their occurrences, on a time scale of hours or minutes. They demonstrate that the P=0.10333 day is not an artifact of some selection effect due to the recording of the flares through the particular *Chandra* observational window (see also Section 5.1 below).

With a finer division of the [0,1) range of the *a* value I find that the most tight grouping of the 39 flares of N13 is found for *a*=0.33. The dispersion of the points marking the first third of each flare around the tic marks of the grid with the period P has a StD value of S=0.16145.

The ephemeris of the grid on the time axis is:

 (4)                    t = 55965.99948+0.1033306*E d

I also find that with *a*=0.33 the best goodness of fit parameter (lowest S value) in the period range [0.02-0.5] d is obtained for P=0.10333 d not only in the entire set of the 39 time points but also in subsets of them. For example the points corresponding to the subsets of flares numbers 16 to 28, 19 to 39, or 11 to 18 combined with 29 to 39, all show tightest grouping around the same P value. For the two disjoint subsets 1 to 24 and 25 to 39, I also find, in each one of them separately, that in the range [0.1-0.5] d the tightest grouping is around P=0.10333 d. These findings may serve as an additional piece of evidence for the non-randomness of the P value.

## 4. Newer Data

### 4.1 The 2012 blocks

Some two years after the appearance of the N13 paper, P15 published the parameters of 92 flares, recorded by *Chandra* between the years 2000 and 2014, and 23 flares that were recorded by *XMM Newton* during a similar time interval. The *Chandra* list includes the observations performed in the 2012 XVP project that were analyzed in the previous sections. The new paper reports on blocks rather than on flares. The algorithm that was used to define a block within the string of data is different from the one used in N13 to define a flare in the data. The new list of the 2012 observations includes 46 entries, as compared with 39 in N13. There are 7 reported flares in the 2013 list that do not have counterparts in the 2015 list. In return, among the 46 blocks of the 2015 list there are 5 that do not have counterparts in the 2013 flare list. There are 14 other blocks judged by the algorithm of 2015 as distinguishable from one another that fall within just 5 time segments that are judged by the 2013 algorithm as distinguishable flares. Also the beginning and the end times of each of the 32 entries that are common to the two lists are slightly different between the two.

Applying the 3D search routine described above on the 46 blocks of P15 I find the same period P among these data with *a*=0. This means that for the set of blocks, the beginning times are statistically the points that are grouped most tightly around tics of an equi-distance grid on the time axis, which has however precisely the same periodicity P=0.10333 d, as found for the 39 flares. The corresponding StD value is S=0.18703

In the P15 list of blocks, numbers 1, 2 and 3 form a single continuous time interval, with no gaps between them, that nearly coincides with flare No. 2 of N13. In a further analysis of the P15 data I consider these 3 blocks as one. Also each one of the following group of blocks [26,27], [35,36,37,38], [40,41,42], [43,44] covers a continuous time interval between the beginning time of the first block in the group and the stop time of the last one. I consider also each of these groups as one block with the beginning time of the first one and the stop time of the last one. This is done very much in accordance with the remark made by P15 that "bright or very bright flares can present significant substructures generating more than one flaring block for each flare". Thus we are left with 37 blocks, as defined by the algorithm used by P15, that are well separated from one another and that no two of them correspond to a single flare of the N13 list.

On this set of 37 blocks I apply again the 3D search routine. The best S value found for this set is again for $a$=0. The p value found is identical to the value P=0.10333, up to the 6th decimal figure. The corresponding StD value is S=0.16549.

To estimate the FAP of the result obtained for the 37 blocks of P15 I performed a numerical simulation on a sample of 37 sets of pseudo-observed blocks, as was done for the 39 flares of N13. From a sample of the 1682 sets I obtain the estimation $FAP_1 = 0.0073807$, corresponding to $2.68\sigma$ significance level and $FAP_2 = 0.011891$, corresponding to $2.52\sigma$ confidence level.

As mentioned above there is a group of 32 flares and a group of 32 blocks that closely coincide with each other. The difference in the statistical significance between the results for the 39 flares and the results for the 37 blocks must therefore reside in the 7 flares that were not recognized as blocks, and the 5 blocks that were not identified as flares. One finds that 6 of these 7 flares are the 6 shortest flares detected, with the 7th one being No. 13 when the 39 flares are listed in an ascending order of their duration. The very short nature of these flares makes them particular suitable for the analysis that we are performing since the grouping of the points that represent their timing is insensitive to the value of $a$ in expression 3 above. These flares are therefore especially good markers for the presence of periodicity among the flares.

Among the 5 blocks for which no counterparts are found in the list of flares, 3 are the blocks with the smallest values of count rate among the 46 blocks of P15. The other 2 are the 7th and the 11th when the blocks are listed in an ascending order of their count rate. Also 3 of these blocks are among the 6 blocks with the longest duration, the 4th one is the 10th in the list when ordered by duration, and the 5th one is the 17th on this list.

The lower statistical significance of the periodicity among the 37 blocks may therefore be understood as resulting from the omission of the 7 best markers of period from this group of blocks on one hand, and from the inclusion of 5 blocks of relatively small signal and probably not well defined structure. It is also possible that these 5 blocks belong to another family of very weak flares, of the kind discussed in section 4 in N13, that indeed do not share the periodic regularity with the class of the more intense ones. In any event, the period P is detected even in this set of 37 blocks at a confidence level of at least 2.5 $\sigma$, as mentioned above.

While the finding of the period P among the 37 blocks is clearly not an independent result from the finding of this periodicity among the 39 flares, it is not an entirely redundant result either. The 5 additional blocks over the 32 that coincide with flares and the 7 flares not included in the set of the 37 blocks, as well as the difference in the beginning and the end times between the blocks and the flares, do affect the results of the periodic grid search procedure, as is evident from the best value *a*=0 found for the set of blocks rather than *a*=0.33 as found for the set of flares. In contrast, the value of the period P=0.10333 d is precisely the same in the two sets. Also, as mentioned above, the very same period is found among the 46 blocks that are on the original P15 list. The robust nature of the period value P may add some further credibility to the claim that the period P is a feature of the Sgr A* object.

## 4.2 Flares before and after 2012

Table 4 in P15 presents 24 flare blocks recorded between the years 2000 and 2011 (one line in this table is printed twice inadvertently). As the mean frequency of these recorded flares is about 2 per year, they constitute a rather poor data base for a search for a periodicity among them of the order of a fraction of a day.

Table 6 in P15 presents data of 22 blocks recoded in the years 2013-2014. Here the density of recorded events is of course higher but still much lower than in the 2012 data. No trace of any periodicity could be found among 2013-2014 data.

## 5. Discussion

## 5.1. Further notes on significance

In order to mimic in my numerical simulation the circumstances under which the real data were collected even further, I considered another sample of pseudo-observed flares created in the following manner. As note by N13 and already mentioned above, the average frequency of X-ray flares of Sgr A* is about one per day (see, for example Markoff (2010), Li et al (2015)). Following N13 who estimated a mean frequency of 1.1 flares per day, I take 290 as a number of flares that took place during the 265 days of *Chandra* XVP project. Accordingly, each one of the sets of the pseudo-observed flares is built on the basis 290 beginning times chosen randomly from a rectangular probability distribution over the time interval of 265 days. To each beginning time I attached a flare length, selected randomly, with repetitions, from the ensemble of the lengths of the real, recorded flares, as derived from Table 1 in N13. I then considered all those flares or fractions of flares that fall within the 38 time slots of *Chandra's* observational window in 2012, also presented in Table 1 of N13. In this way I created 11714 sets of pseudo observed flares. On each of these sets I apply the 3D search routine as done for the real data. In the sample of 11714 sets of pseudo observed flares so created I find just 2 sets with $S \leq 0.16222$ and $p \geq 0.10333$. The false alarm probability is therefore 2/11714=0.00017, corresponding to significance level of 3.7$\sigma$.

According to this test, the significance of the finding in the real data seems to be even much higher than implied by the previous 2 tests. However, I believe that not too much weight should be given to the estimate made with this one. The statistical space underlying this test includes also all possible events where within the 38 time slots of the observational window of *Chandra* XVP project (see N13), not only 39 but also different number of flares could have been recorded. The a-priori probability of *Chandra* to record just 39 flares should not be a part in our considerations. In this study we are investigating whether or not there is a statistical regularity in the times of flares, within the day of their occurrence. The time scale of our concern is not of days but of the order of hours or minutes.

## 5.2. Comment on the validity of the numerical tests

The search routine and the tests of the statistical significance of its finding that was applied in this work are seemingly similar to time series analysis that was performed in the context of other research fields. For example, Raup and Sepkoski (1984 hereafter - RS) claimed a discovery of a period of 26 Myr in the occurrence of peaks in the rate of species mass extinction in the geological history of Earth. The technique used by these authors in unveiling this periodicity in the time series of extinction rates measured in 39 stages along the past 250 Myr of Earth history, has some resemblance to the analysis made in this work. The conclusion of RS was criticized by Stigler & Wagner (1987 - SW), Bailer-Jones (2009) and others. SW, for example, have agreed that the simulation test performed by RS permits the rejection of the null hypothesis that the apparent periodicity reflects a random process, at the significant level of the test. They have argued, however, that the rejection of this null hypothesis does not entail periodicity. SW have shown examples, where application of the analysis of RS on time series known to be non periodic, may nevertheless yield periodicity as an outcome.

There is, however, a fundamental difference between the RS approach and the search routine applied in this work. For one thing, we are dealing here with 39 time points as compared with 12 or even merely 8 points analyzed by RS. These authors were looking for, and apparently finding, a periodicity in a measured variable R, the rate of extinction of species on Earth, which is presented as a dependent parameter on the (discrete) time coordinate t, an independent one. Accordingly, the random element in their simulations, as well as in those of SW, is introduced in the distribution of the dependent parameter R, while the time coordinates are held fix at the times of the 39 geological stages considered by RS. In contrast, there is no dependent variable in the time series of our interest. Once the selection of X-ray flares were made by the algorithm used by N13, and as we have seen, also by P15, the periodicity that we are finding is related (through the fitting of the grid on the time axis) directly to the timing of the events. Accordingly, our simulations created pseudo-observed timings of flare rather than simulating different distributions of some parameter that is a function of these times. As an illustration of the difference between the treatment of RS in their time series analysis and the one performed in this work one may compare Figure 4b of this work with Figure 4 in the paper of SW. As explained by SW, their figure demonstrates that the apparent periodicity in the peaks of extinction rates found by RS is related to the uneven distribution, or lengths, of the geological stages at the basis of the time series analyzed by RS. In our case, Figure 4b shows that the P=0.10333 d uncovered by our analysis is not hidden in any ways in the distribution of other parameters that characterize the flares, such as their length.

## 5.3. Suggested qualitative interpretation

It is by now generally accepted that the X-ray and IR flares of Sgr A* emanate from the very near vicinity of the object at the center, from distances that are of the order of the gravitational radius of the BH. For example, Genzel et al (2003) report on 17 minutes quasi-periodic variations in the IR radiation from within a few milli-arcseconds of the BH. They suggest that this emission comes from just outside the event horizon. Ghez et al (2004) report also on varying IR radiation from Sgr A*, the source of which they place no further than 5 AU, about 80 Schwarzchild radii from the central object. These authors suggested that the same source is shared also by the X-ray flares of the object. Doeleman et al (2008) write that activity on scales of a few Schwarzscild radii in Sgr A* is responsible for observed short-term variability and flaring from radio to X-rays. They suggest that the varying component in the radiation is excited by MHD turbulence, and the excess power is associated with inward propagating magnetic filaments inside the innermost stable circular orbits (ISCO) of the BH.

Recently, Alston et al (2015) report on quasi-periodic oscillations in the X-ray radiation from the active nuclei of 2 Seyfert galaxies with a period of about 1 hour. They attribute the oscillations that they discovered to the accretion process onto the BHs at the centers of these galaxies.

Wang et al (2014) find some statistical similarities between solar flares and X-ray flares in Sgr A* and in AGNs. They suggest that the flares in these two types of objects share a common origin in being consistent with magnetic reconnection events.

X-ray quasi-periodic variability associated with processes that take place very near the event horizon of a SMBH, not only at the center of the Galaxy but also in external galaxies, may therefore be ubiquitous. The new element in the discovery of the periodic element in the X-ray emission from the center of the MW is that it relates to the occurrence of flares, rather than to a continuous emission process. It indicates an operation of an agent with a strict, rather than quasi, periodicity, and that this periodicity endures in the data for at least 9 months.

The characteristic time of the QPO in the Sgr A* IR and X-ray radiation is of the order of a few, or a few tens of minutes which is the dynamical scale of events in close proximity to the ISCO. The larger value of the $P \cong 150\ min$ and its coherence over 9 months make turbulent MHD processes an unlikely source for the timing of the flares. A natural candidate clock in the system of coherent frequency could be a Keplerian orbital revolution.

In fact, Broderick and Loeb (2005) suggested already some 11 years ago that the agent responsible for the X-ray flares of Sgr A* could be an object in circular orbit around the BH. These authors were mainly concerned with the effect of the curving of space near a rotating BH and their computations were performed in the frame of the Kerr metric at distances from the center of the order of the ISCO. However, as pointed out by Abramowicz and Fragile (2013), around a BH, Keplerian circular orbits exist in the region $r > r_{ph}$, with $r_{ph}$ being the circular photon orbit, and stable orbits exist in the region $r > r_{ms}$, with $r_{ms}$ being the radius of the ISCO. If the period that we discovered among the recorded flares in 2012 is the period of one such Keplerian orbit, then for $M_{BH} = 4.4 \times 10^6 M(Sun)$ (Genzel, Eisehauer and Gillessen, 2010), and taking into account the gravitational time dilation, we find the radius of this orbit to be $7.6 R_s = 9.9 \times 10^7 km = 0.66 AU$, where $R_s$ is the Schwarzschild radius of the BH. The orbital velocity is 0.26c, where c is the velocity of light.

The lack of any sign for periodicity in the years 2013-2014 may simply be due to the still rather meager observational data for revealing the presence of a period of a similar value as the one found in the 2012 data. It may, however, be related to the finding of P15 that over the 2013 and 2014 period, the total energy emitted by Sgr A* in the form of bright-or-very bright flares, increased by factor 3.7 compared to historical observations. Thus it might be that this increased rate overwhelmed the regularity imposed on the generation of observed flares as reflected by the presence of the periodicity in the year 2012. Alternatively, we should infer that the pacemaker mechanism in the Sgr A* system, at least with a period of the order of a fraction of a day, is in general a transient phenomenon with a duration time of the order of a few months.

One possible interpretation of the result of this work could be that the orbit of the agent responsible for the generation of outbursts is not strictly circular. Outbursts of large flares are more likely to be triggered when this agent is closer to its peri-center point of its orbit around the BH. We may further speculate that this agent is an M dwarf star of mass $m_d = 0.3 M(Sun)$. In a non relativistic approximation we can estimate the radius of the equi-potential Roche lobe $r_L$ of such a star as a binary companion of the SMBH, using equation (2) of Eggleton (1983). For the mass ratio $q = \frac{0.3}{4.4 \times 10^6} = 6.8 \times 10^{-8}$ and the binary separation of $7.6 R_s$ we find $r_L = 0.28 R(Sun)$. According to the mass-radius relation of low mass stars presented in Figure 8 by Demory et al (2008), the radius of a star of mass $m_d$ is just about $r_L$.

We may therefore speculate that the agent responsible for the large X-ray flares in the year 2012 was an M dwarf star, revolving around the BH in a slightly eccentric Keplerian orbit. The configuration was such that the star was just filling its equi-potential Roche lobe. At peri-center passages, the radius of the Roche lobe of the star is slightly shrinking. Large episodes of mass accretion from this stars were therefore more likely to occur at orbital phases around peri-center passages, giving rise to the observed large X-ray flares. We note, however, that this scenario leaves quite open the question of what is the future fate of this star and how it will be reflected in the luminosity of Sgr A*.

## 6. Summary

The 9 months long *Chandra* XVP project was a unique, rare experiment revealing a peculiar statistical feature in the distribution in time of the 39 X-ray flares that have been recorded by this telescope during that period. The time points, each one marking the end of the first third of its corresponding flare duration, possess a very special statistical quality. Within a rather small dispersion in the [-0.5,0.5] phase interval of StD=0.16145, they are grouped around tic marks of equi-distance grid on the time axis of the periodicity P=148.8 min. This finding can be accepted at a 3.2 $\sigma$ confidence level ($FAP \cong 1/700$) as not being a result of a random coincidence but rather as a genuine feature of the X-ray flare phenomenon of Sgr A* in 2012.

Following others we suggest that P is the period of a Keplerian orbit around the BH, and if so, its radius is about 7.6 Schwarzschild radii of the BH. The orbit, however, is not strictly circular. Outbursts of large flares are more likely to be triggered when the agent responsible for their generation is closer to its peri-center point of its orbit around the BH. This agent could be an M star of mass $0.3R(Sun)$.

The time point sets that were analyzed in this work were defined by the beginning and end times of the flares. It may be worthwhile to scrutinize carefully the detailed structure of each of the 39 flares, in particular in sections around one third of their duration. There, one may look for a possible morphological sign of the exact time points that are modulated by the so far unknown periodic process in the system.

## Acknowledgement

I am indebted to an unknown referee for meticulously reviewing this paper and for improving it by suggesting very useful additions and corrections.